\documentclass{elsart}
\usepackage{graphicx}
\usepackage{amssymb}

\def\R{\mathbb R}
\begin{document}
\begin{frontmatter}
\title{Modeling the Heating of Biological Tissue
       based on the Hyperbolic Heat Transfer Equation}

\author[mike]{M.\ M.~Tung\corauthref{cor}}\ead{mtung@imm.upv.es},
\corauth[cor]{corresponding author}
\author[impa]{M.~Trujillo}\ead{matrugui@mat.upv.es},
\author[impa]{J.A.~L\'opez Molina}\ead{jalopez@mat.upv.es},
\author[impa]{M.J.~Rivera}\ead{mjrivera@mat.upv.es}, and
\author[enrique]{E.J.~Berjano}\ead{eberjano@eln.upv.es}

\address[mike]{Instituto de Matem\'atica Multidisciplinar}
\address[impa]{Instituto de Matem\'atica Pura y Aplicada}
\address[enrique]{Instituto de Investigaci\'on e Innovaci\'on en Bioingenier\'{\i}a
\\[.4cm] Universidad Polit\'ecnica de Valencia, Valencia, Spain}

\thanks{This work received financial support from the Spanish
``Plan Nacional de Investigaci\'on Cient{\'\i}fica, Desarrollo e
Innovaci\'on Tecnol\'ogica del Ministerio de Educaci\'on y Ciencia''
(TEC 2005-04199/TCM) and from the MEC and FEDER Project MTM2007-64222.
}

%\maketitle

\begin{abstract}
In modern surgery, a multitude of minimally intrusive operational
techniques are used which are based on the punctual heating
of target zones of human tissue via laser or radio-frequency
currents. Traditionally, these processes are modeled by the bioheat
equation introduced by Pennes, who considers Fourier's theory of
heat conduction. We present an alternative and more realistic
model established by the hyperbolic equation of heat transfer.
To demonstrate some features and advantages of our proposed
method, we apply the obtained results to different types of
tissue heating with high energy fluxes, in particular radiofrequency
heating and pulsed laser treatment of the cornea to correct refractive errors.
Hopefully, the results of our approach help to refine surgical
interventions in this novel field of medical treatment.
\end{abstract}

\end{frontmatter}

% main text

\section{INTRODUCTION}
\label{intro}

Presently, modern surgery has at its disposal a great variety of different
surgical techniques to heat biological tissue in a localized and safe way.
All these techniques are based on specific applicators, {\it i.e.}\/
devices which extract (cryosurgery) or introduce heat 
(laser, radiofrequency current, microwave or ultrasound treatments). 
Theoretical heat modeling is a very cheap and fast methodology to study the
thermal performance of these applicators. In fact, a lot of previous
work has been conducted to model these procedures by using the bioheat
equation as governing equation (see \cite{Berjano}, and references therein).
This equation is based on the classical Fourier theory of heat conduction
and is widely used for modeling the heating of biological tissue.
Nevertheless, it has been hypothesized that for heat transfers on very small time
scales the classical model will fail, and an alternative thermal wave theory
with a finite thermal propagation speed could be more suitable to
describe these phenomena.

For the aforementioned reasons, we are lead to modeling the surgical
heating of biological tissue by means of the hyperbolic heat transfer equation,
and then compare these alternative results with those obtained from the
standard Fourier theory (implying the parabolic heat transfer equation).

In summary, the main theme of this work is to present in depth the mathematical
and physical background for a general discussion of heat models related to heating
of biological tissue by means of energy applicators, and then, in
particular, point out the physical differences between the classical
Fourier theory and the hyperbolic wave theory. Moreover,
as an illustrative example, we give some numerical results of the analytical
modeling for radiofrequency heating ({\sl RFH}\/) and for laser heating, each
applied to the cornea in order to correct refractive errors.
Both surgical techniques have in common that they may involve
high heat transfers during very short exposure times.

This paper is organized as follows.
In Sec.~\ref{heatmodels} we will set up the mathematical groundwork for 
heat models related to {\sl RFH} and laser heating with a special emphasis
on parabolic and hyperbolic heat models.
Then, in Sec.~\ref{applications} we present the results for typical
laser and {\sl RFH} interventions in order to substantiate the physical
discrepancies between both heat models in concrete examples. We will
also comment on the domain of applicability of both theories and their
main features.
Finally, Sec.~\ref{conclusions} will present the conclusions and give
an outlook on interesting future work.

\section{HYPERBOLIC VERSUS PARABOLIC HEAT MODELS}
\label{heatmodels}
\begin{figure}[h]
\centering
\includegraphics{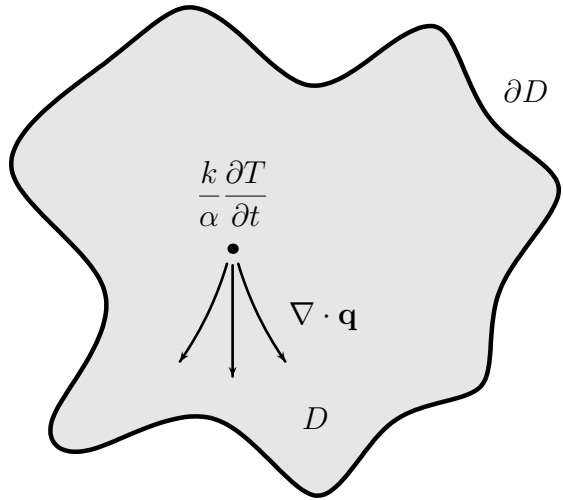}
\caption{Schematic view of thermal current conservation. The heat source
$S(\mathbf x,t)$ in a physical system ($D\in\R^3$ simply-connected
domain) is related to both, temperature change and divergence of the
heat flux $\mathbf q$. The coefficient $k/\alpha$ gives the volumetric
heat capacity in terms of thermal conductivity and diffusivity.}
\label{currentpic}
\end{figure}

The most fundamental relation to model heat transfer is the generally
valid, {\it i.e.\/} model-independent, equation for thermal current
conservation with a given internal heat source $S(\mathbf x,t)$:
\begin{equation}
\label{current}
      \nabla\cdot\mathbf q(\mathbf x,t) +
      {k\over\alpha}{\partial T\over\partial t}(\mathbf x,t)
      = S(\mathbf x,t).
\end{equation}
Here, $\mathbf q(\mathbf x, t)$ denotes the thermal flux
and $T(\mathbf x, t)$ is the temperature at point
$\mathbf x\in D$ in the domain $D\subset\R^3$ at time $t\in\R_+$.
As usual, thermal conductivity is denoted by $k>0$ and diffusivity
by $\alpha=k/\rho c$, where $\rho c$ is the volumetric heat capacity,
being $\rho$ the density and $c$ the specific heat of the material
under consideration.
Fig.~\ref{currentpic} gives a schematic view of thermal current conservation.

\begin{figure}[h]
\centering
\includegraphics{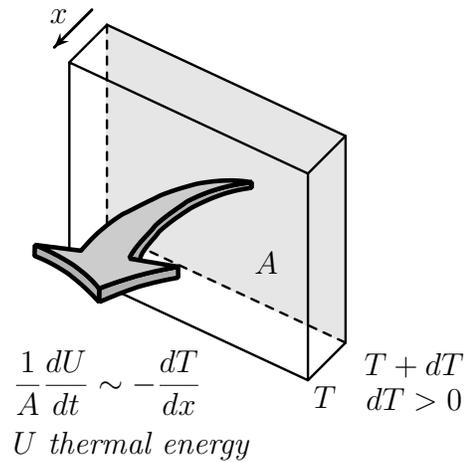}
\caption{Fourier's law in differential form, shown for heat transfer through
an infinitesimal thin plate with thickness $dx$ and cross-sectional surface area $A$.
The corresponding flux, thermal energy per time and area, is proportional to the
temperature gradient $dT/dx$. The sign indicates that heat flows from $T+dT$ to $dT$.}
\label{fourierpic}
\end{figure}

Fourier's law of heat conduction \cite{Fourier} constitutes the foundation of all classical heat models.
It proposes that the heat flux $\mathbf q(\mathbf x, t)$ is proportional to the negative
of the temperature gradient (see Fig.~\ref{fourierpic}):
\begin{equation}
\label{fourier}
\mathbf q(\mathbf x, t) = -k\,\nabla T(\mathbf x, t).
\end{equation}
However, in this relation temporal changes instantaneously affect heat flux and
temperature gradient. As an immediate consequence any perturbations in classical
heat models are propagated with infinite speed.

Assuming that thermal current conservation (\ref{current}) and Fourier's law (\ref{fourier})
hold, directly yields the classical, {\it i.e.\/} parabolic heat transfer equation
({\sl PHTE}\/) with heat sources:
\begin{equation}
\label{PHTE}
      -\Delta T(\mathbf x,t)+{1\over\alpha}{\partial T\over\partial t}(\mathbf x,t)
      = {1\over k} S(\mathbf x,t).
\end{equation}
The standard bioheat equation was introduced by Pennes \cite{Pennes} and is based on
the parabolic heat transfer model, identifying the following explicit contributions
for the heat sources in a biological system
\begin{equation}
\label{bioheat}        
      S(\mathbf x,t) = 
      S_s(\mathbf x,t) +
      S_p(\mathbf x,t) +
      S_m(\mathbf x,t),
\end{equation}
where the subindex $s$ denotes a surgical heat source ({\it e.g.\/} laser or radiofrequency
treatment), $p$ refers to blood perfusion, and $m$ to any source related to metabolic activity.

With the advancement of modern surgery, medical treatments involve
progressively smaller time scales and higher energy fluxes. For example,
for radiofrequency and laser surgery {\sl PHTE} models could become
inappropriate. The reason is that on small time scales $t\in[0,\tau]$, with
a sufficiently small value for $\tau>0$, thermal equilibrium of an
extended physical system simply can not be reached. This contradicts the
classical model, where all elements of the thermodynamic system interact
with no delay through infinite thermal speed. It is to be expected that
such theories implying infinite speed of perturbations in the heated
medium and more realistic wave theories with finite speeds differ
considerably in their predictions.

A modified version of Fourier's law (\ref{fourier}) uses a non-vanishing
relaxation time $\tau$ in the dissipative process. This parameter may be
interpreted as the finite time necessary for the dissipative flow to relax to
its steady thermodynamic value.
The simplest generalization for finite speeds leads to Cattaneo-Vernotte's
equation \cite{Cattaneo,Vernotte}:
\begin{equation}
\mathbf q(\mathbf x, t+\tau) = -k\,\nabla T(\mathbf x, t).
\end{equation}
Note that the classical Fourier law is still included as a special limiting case,
taking the limit of zero relaxation time for the heat fluxes. In
general, however, there is an inertial or retardation term which delays
any changes in the temperature gradient to be transferred to the heat
flux. This can also be seen for small relaxation intervals $\tau$, which
yields the expression
\begin{equation}
\label{cv}
      \mathbf q(\mathbf x, t) +
      \tau {\partial\mathbf q\over\partial t}(\mathbf x, t)
      = -k\,\nabla T(\mathbf x, t).
\end{equation}
Although it appears at first sight that Eq.~(\ref{cv}) is an expansion
which only holds approximately for small relaxation intervals $\tau$,
by considering entropy arguments \"Ozi\c{s}ik and Tzou have shown
that Eq.~(\ref{cv}) is also valid on a macroscopic scale~\cite{Ozisik}.

By combining the heat transfer model (\ref{cv}) with current
conservation (\ref{current}), one directly obtains the hyperbolic heat
transfer equation ({\sl HHTE}\/) with heat sources:
\begin{equation}
\label{HHTE}
      -\Delta T(\mathbf x,t)+{1\over\alpha}\left(
      {\partial T\over\partial t}(\mathbf x,t)
      +\tau{\partial^2 T\over\partial t^2}(\mathbf x,t)\right)
      = {1\over k}\left(
      S(\mathbf x,t)+
      \tau{\partial S\over\partial t}(\mathbf x,t)\right).
\end{equation}
For zero relaxation the
hyperbolic heat equation obviously reduces to the classical model.
An essential feature of Eq.~(\ref{HHTE}) is the second-order time derivative
with coefficient $\tau>0$, which gives rise to wave properties of the equation,
and therefore implies finite thermal propagation speeds. This can easily
be seen by considering forwardly propagating waves which provide solutions
for Eq.~(\ref{HHTE}) of the form
\begin{equation}
\label{wave}
T_{\!\hbox{\scriptsize wave}} :=
e^{i\left(\mathbf k\cdot\mathbf x-\omega t\right)}f(\mathbf x-\mathbf v t),
\end{equation}
where $\mathbf k$ is the usual wave propagation vector and $\omega$ the
corresponding frequency.
This approach is somehow analogous to the analysis of the {\it telegraph equation},
which describes the propagation of electromagnetic waves in conducting
media~\cite{Sommerfeld,Panofsky}. Note that in the present case, however, a term
proportional to $T(\mathbf x,t)$ is absent on the left-hand side
of Eq.~(\ref{HHTE}).

By following the procedure of Ref.~\cite{Sommerfeld} and substituting
Eq.~(\ref{wave}) into Eq.~(\ref{HHTE}) for sourceless heat transfer,
then requiring that the real-numbered factor multiplying with
$\Delta f$ has to be zero, one directly obtains for the thermal
propagation speed 
\begin{equation}
\label{speed}
v=\sqrt{\alpha\over\tau}\quad\hbox{with}\quad \tau>0.
\end{equation}
From this it becomes evident that for vanishing relaxation times $\tau$ and
constant diffusivity $\alpha>0$, the propagation speed tends to infinity.
As indicated before, but seen from a different perspective, in this limiting
case the hyperbolic model reduces again to the standard Fourier theory.

One further distinctive feature of the {\sl HHTE} model is observed
by representing the hyperbolic heat flux equation (\ref{cv}) in integral
form~\cite{Ozisik}
\begin{equation}
\label{hypflux}
\mathbf q(\mathbf x,t) = -{k\over\tau}\,e^{-t/\tau}
\int\limits_0^t e^{s/\tau}\,\nabla T(\mathbf x,s)\,ds.
\end{equation}
Note that Eq.~(\ref{hypflux}) shows that $\mathbf q(\mathbf x,t)$ depends on the full
history of the temperature gradient within the time interval $[0,t]$, something
which is entirely absent in the much simpler expression for the Fourier flux
in Eq.~(\ref{fourier}).

Exhibiting these features, the hyperbolic heat transfer model with
its finite thermal propagation appears to provide an ideal framework
to describe surgical interventions such as laser ablation
(a procedure named as laser thermokeratoplasty, or short {\sl LTK}\/)
and {\sl RFH}\/ of the cornea ({\sl CK}, conductive keratoplasty), which
use very short and high-energetic pulsations (typical time periods are
$200~\mu\hbox{s}$ and $50~\mu\hbox{s}$, respectively).
Therefore, in the following section, we will focus on applications of heat models
in surgical techniques, in particular {\sl RFH} and laser
surgery applied to the cornea, and thereby develop a realistic model for the
heating of biological tissue by employing the {\sl HHTE} as previously
derived in Eq.~(\ref{HHTE}).

\section{SURGICAL APPLICATIONS IN HEAT MODELS}
\label{applications}

There exist numerous processes in which great amounts of heat are
applied to materials in very short exposure times.
This section will focus on the modeling of heating biological tissue as it
occurs during medical intervention with surgical radiofrequency or laser
devices to the cornea.

Laser heating is a process which implies tissue heating caused by absorbing
the optical energy of a high-energy laser beam ({\sl LTK}\/).
On the other hand, radiofrequency heating ({\sl RFH}\/) is ultimately based on
Ohm's law, which states that currents flowing through a resistor
generate heat. It is essentially ionic motion in the tissue which will
provoke biological heating aimed at producing any medical effects.
Both are surgical techniques which may not only involve small time
scales but also high energy fluxes, which makes them an important group
of applications in which differences between parabolic and hyperbolic
models could have great effects.

As we have previously mentioned in Sec.~\ref{heatmodels}, these type
of processes represent non-equilibrium processes on small time scales
({\it i.e.} compared to the relaxation time $\tau$),
since the system requires considerably more time to reach the equilibrium
state after the initial thermal energy input. For this reason, {\sl PHTE}
models, with their infinite thermal propagation speed, may not provide
an appropriate description for the underlying physical structure, and
it may be necessary to rely on {\sl HHTE} models instead.

\subsection{Laser Heating}
\label{laser}

Among the many surgical procedures in which laser heating is
employed, we focus our attention on corneal laser heating,
also referred to as laser thermokeratoplasty ({\sl LTK}\/)~\cite{manns}.

\begin{figure}[h]
\centering
\includegraphics{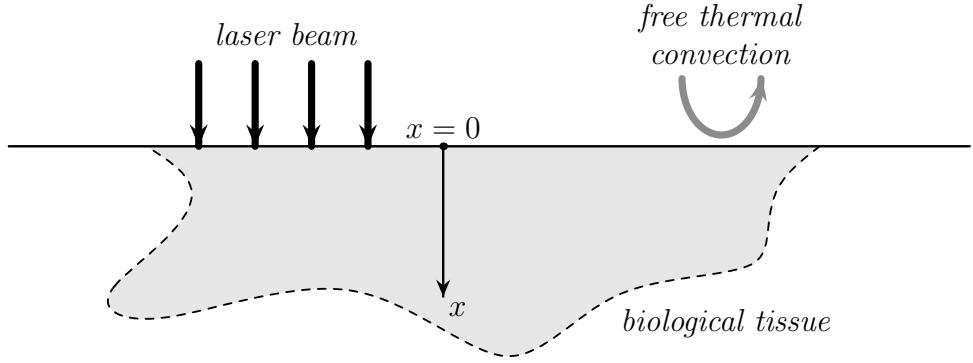}
\caption{A schematic view of the model geometry for laser ablation. The
laser beam is applied perpendicular to the tissue and defines the
$x$-axis. The tissue is assumed to take the shape of the half-plane $x\ge0$
with free thermal convection above.}
\label{laserpic}
\end{figure}

In the case of laser heating, we consider a theoretical model
consisting of a semi-infinite fragment of homogeneous isotropic
biological tissue in which the laser beam falls on the entire
tissue surface. Fig.~\ref{laserpic} depicts the chosen model
geometry, in which we consider a one-dimensional model with
the $x$-axis parallel to the direction of the incident laser beam.
Above the surface, for $x<0$, free thermal convection is supposed
to cool the tissue surface.

As we want to study the problem from the point of view of the
{\sl HHTE} model, we have to start off from the governing equation
provided by Eq.~(\ref{HHTE}). For this one-dimensional laser heating
model, we have to solve the corresponding heat transfer equation
for penetration depth $x$ including the appropriate heat-source
term and boundary conditions.

Neglecting blood perfusion and metabolic activity in Eq.~(\ref{bioheat}),
and hence $S_p=S_m=0$, the heat source $S(x,t)$ will only have
contributions from $S_s(x,t)$. This surgical source should be obtained
from the Beer-Lambert law, which empirically states that for
radiation the intensity decreases exponentially with penetration
depth. Including a factor with the temporal dependence to model
an energy pulse of duration $\Delta t$, this yields
\begin{equation}
\label{beer}
S(x,t) = (1-R)\,b\,E_0\,e^{-bx}\Big[H(t)-H(t-\Delta t)\Big],
\end{equation}
where $R$ denotes the dimensionless Fresnel surface reflectance,
$b$ is the absorption coefficient, and $E_0$ is the incident energy flux
at the tissue surface~\cite{manns}. As usual, $H(t)$ denotes the
Heaviside function. After combining Eqs.~(\ref{HHTE}) and (\ref{beer}),
the final form of the governing equation is
\begin{equation}
\begin{array}{l}
\displaystyle{
-{\partial^2T\over\partial x^2}(x,t)+
{1\over\alpha}\left({\partial T\over\partial t}(x,t)
+\tau{\partial^2 T\over\partial t^2}(x,t)\right)}
= \\
\displaystyle{
(1-R){b E_0\over k}e^{-bx}
\Big[H(t)-H(t-\Delta t)+\tau\left(\delta(t)-\delta(t-\Delta t)\right)\Big]},
\end{array}
\end{equation}
where $\delta(t)$ is the Dirac delta function.

The corresponding initial boundary conditions are given by
\begin{equation}
\begin{array}{lcc}
 T(x,0)=T_0, & \hspace{-2cm}\displaystyle{\partial T\over\partial t}(x,0)=0 &
    \qquad\forall x>0 \\[.25cm]
 \lim\limits_{x\to\infty} T(x,t)=T_0 & & \qquad\forall t>0 \\[.25cm]
 \displaystyle{\partial T\over\partial x}(0,t) = 
 \displaystyle{
 {h\over k}\left(\tau \frac{\partial T}{\partial t}(0,t)+T(0,t)-T_a\right)}
    & & \qquad\forall t>0
 \label{coolflux}
\end{array}
\end{equation}
where $T_0$ is the initial temperature and $T_a$ is the ambient temperature.
Furthermore, $h>0$ is the thermal convection constant at the interface tissue
given by Newton's law of cooling, which states that the convective flux of
an object is proportional to the difference between its own temperature $T_0$
and the ambient temperature $T_a$:
\begin{equation}
  q(0,t) = h \left[T_a-T(0,t)\right].
\end{equation}
Since at the interface tissue with $x=0$ the heat flux is given by the
Newton's law of cooling, the last condition of Eq.~(\ref{coolflux}) is
obtained by imposing this law onto the relation for the hyperbolic heat flux
Eq.~(\ref{hypflux}) (see also \cite{Ozisik}).

The explicit analytical solution of this problem has been obtained in
Ref.~\cite{sinpublicar} and is mainly based on the use of
Laplace transforms. It also contains for comparison the fully analytical
{\sl PHTE} solution as an application to the thermokeratoplasty technique.
For the numerical estimates, well-established thermal and optical properties
of the cornea were used. All essential physical constants for the model
are summarized in Tab.~\ref{corneaparam}.

\begin{table}
\label{corneaparam}
\centering
\begin{tabular}{|c|c|c|c|}
\hline
\multicolumn{4}{|c|}{thermal properties} \\
\hline
  density & conductivity & diffusivity & convection \\
  $\rho$ [kg\kern1pt m$^{-3}$] & $k$ [W\kern1pt m$^{-1}$\kern1pt K$^{-1}$] & $\alpha$ [m$^2$\kern1pt s$^{-1}$] & $h$ [W\kern1pt m$^{-2}$\kern1pt K$^{-1}$] \\
\hline
  $1060$ & $0.556$ & $1.3695\cdot10^{-7}$ & 20 \\
\hline
\end{tabular}        
\begin{tabular}{|c|c|}
\hline
\multicolumn{2}{|c|}{optical properties} \\
\hline
  Fresnel reflectance & absorption coefficient \\
  $R$ & $b$ [m$^{-1}$] \\
\hline
  0.024 & 2000 \\
\hline
\end{tabular}
\vskip.35cm
\caption{Thermal and optical properties of the human cornea.}
\end{table}

\begin{figure}[h]
\centering
\includegraphics[width=13cm]{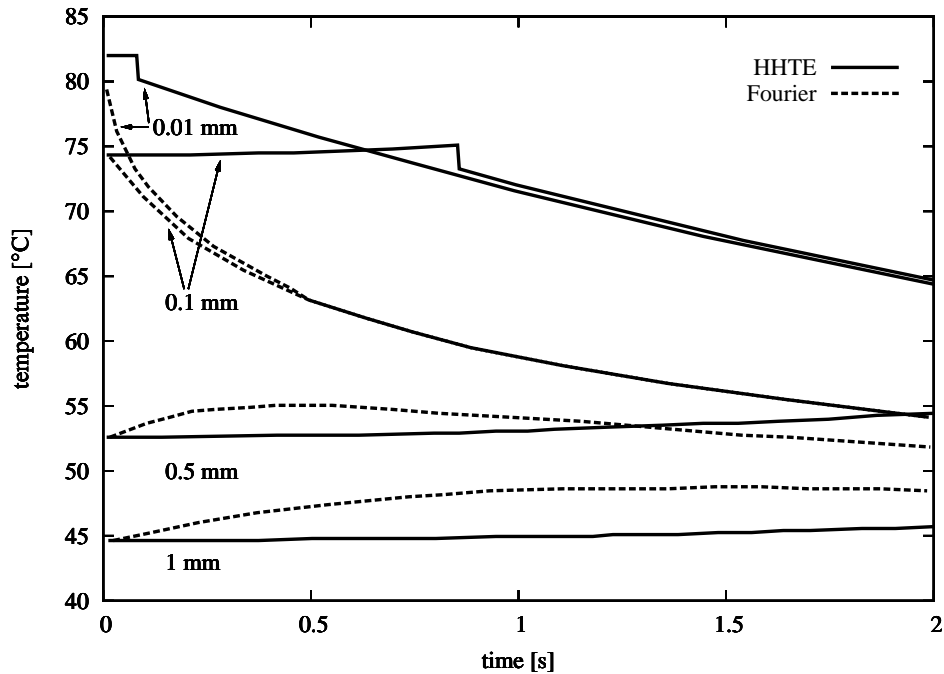}
\caption{Temperature profiles for laser heating of the cornea for an
initial temperature $T_0=35^{\circ}\hbox{C}$ and ambient temperature
$T_a=20^{\circ}\hbox{C}$. The laser pulse has a duration of
$\Delta t=200~\mu\hbox{s}$ with an incident energy flux
$E_0=5\cdot10^8~\hbox{W/m}^2$.
The represented curves for the {\sl HHTE} (solid lines) and Fourier (dashed lines)
model correspond to penetration depths $x=0.01, 0.1, 0.5, 1~\hbox{mm}$.
The relaxation time is $\tau=10~\hbox{s}$.
}
\label{laserabl}
\end{figure}

Figure~\ref{laserabl} shows the numerical estimates for temperatures at
various tissue penetration depths obtained as a function of time~\cite{sinpublicar}.
This figure represents the hyperbolic (solid line) and parabolic (dashed
line) temperature evolution at the four locations $x = 0.01, 0.1, 0.5$ and
$1$~mm for the thermokeratoplasty technique.

We can mainly find two differences between both solutions:

Firstly,
according to the parabolic model, for the values $x=0.01$ and $0.1$~mm
the maximum temperature has already been reached and is steadily
decreasing. For $x=0.5$ and $1$~mm, the maximum temperature has not yet
been reached and is initially increasing. For the {\sl HHTE} predictions,
we observe that for all penetration depths in some part of the plot
(or the entire plot) the temperature is increasing. 
The great differences in temperatures
between both models (overall at positions closer to the surface
$x=0.01$ and $0.1$~mm) is even more important when the energy
application is pulsed, since in every new pulse the difference
between the initial temperature could become greater and greater.

Secondly,
we can confirm that the behavior of both solutions is very different.
At values $x=0.01$ and $0.1$~mm, we notice a delay in the abrupt
temperature drop associated with switching off the laser beam. This
delay is related to the finite thermal propagation speed in the
{\sl HHTE} model. Using in Eq.~(\ref{speed}) the numerical value for
diffusivity ({\it viz.}\/ Tab.~\ref{corneaparam}) and for the relaxation time
$\tau=10~\hbox{s}$ predicts
\begin{equation}
        v = 0.117~{\hbox{mm}\over\hbox{s}}
\end{equation}
for the speed of these steps, which exactly agrees with the known finite
speed of the thermal wave in the cornea \cite{sinpublicar}.
This fact demonstrates the wave character of the {\sl HHTE} model.
For depths $x=0.5$ and $1$~mm, however, these drops are not observed,
because they occur outside the range of the time interval considered.

\subsection{Radiofrequency Heating}
\label{RFH}

Radiofrequency heating ({\sl RFH}\/) is a surgical procedure broadly employed
in many clinical areas such as the elimination of cardiac arrhythmias, the
destruction of tumors, the treatment of gastroesophageal reflux disease
and the heating of the cornea for refractive surgery~\cite{otroberjano}.
In the remainder of this section we will focus on a model of the corneal
{\sl RFH} treatment.

\begin{figure}
\centering
\includegraphics{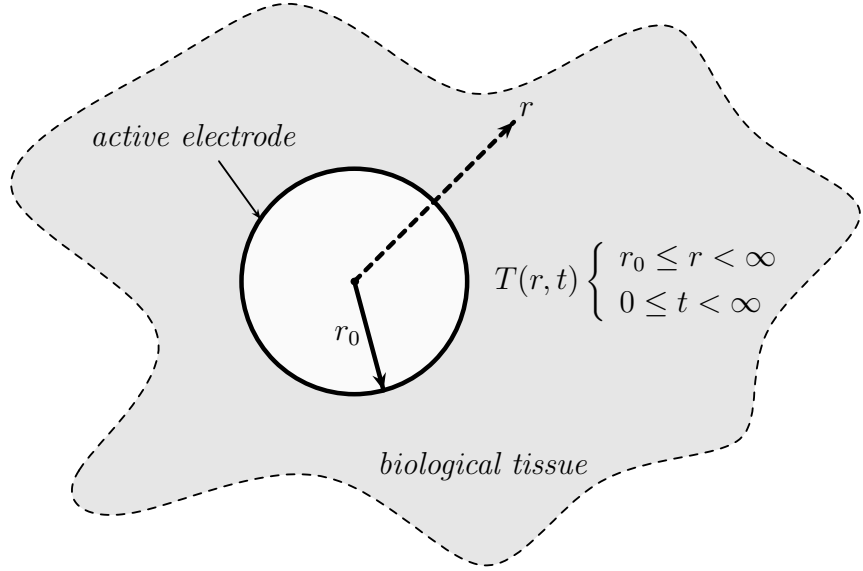}
\caption{A graphical sketch of the model geometry for radiofrequency
heating, showing the intersection for a spherical electrode. The tip
of the active electrode is represented by a sphere of radius $r_0$,
and the biological tissue extends to infinity. Because of radial symmetry, the problem
consists of finding $T(r,t)$, imposing specific boundary conditions.}
\label{modelpic}
\end{figure}

The schematic diagram of the model geometry is shown in Fig.~\ref{modelpic}.
Here, a spherical electrode of radius $r_0$ is completely embedded and in close
contact with the biological
tissue, which has infinite dimension. In this case, we can also
use a one-dimensional model ($r$ being the spatial variable), since
the geometry under consideration displays radial symmetry.

Again, assuming $S_p=S_m=0$ in the bioheat equation (\ref{bioheat}),
the governing equation results from the combination of
Eq.~(\ref{HHTE}) and the explicit form of the (surgical)
heat source $S(r,t)$
which in the case of {\sl RFH} is a product of a radial~\cite{erez}
and a temporal part
\begin{equation}
S(r,t)=S(r)\,H(t)={Pr_0\over4\pi r^4} \,H(t),
\end{equation}
so that the total applied power outside of the electrode is
\begin{equation}
P = \int\limits_{\hbox{\scriptsize\it shell}\ r\ge r_0}
      \kern-10ptS(r)\,dV
      = \oint d\Omega\int\limits_{r_0}^{\infty} r^2 S(r)\,dr,
\end{equation}
where $\Omega$ is the usual solid angle subtended by a surface.
Thus, the power (flux) per unit solid angle, $dP/d\Omega$, satisfies
the usual $1/r^2$ intensity law.
Note that $H(t)$ is included to model a non-pulsed source.
As a result, for the corneal {\sl RFH} the governing equation and
the boundary conditions are:
$$
-{1\over r^2}{\partial\ \over\partial r}\left(r^2
\,{\partial T\over\partial r}(r,t)\right)
+
{1\over\alpha}\left({\partial T\over\partial t}(r,t)
+\tau{\partial^2 T\over\partial t^2}(r,t)\right)
= {Pr_0\over4\pi k r^4}
\Big[H(t)+\tau\delta(t)\Big]
$$
\begin{equation}
\begin{array}{lcc}
 T(r,0)=T_0, & \hspace{-2cm}\displaystyle{\partial T\over\partial t}(r,0)=0 &
    \qquad\forall r>r_0 \\[.25cm]
 \lim\limits_{r\to\infty} T(r,t)=T_0 & & \qquad\forall t>0 \\[.25cm]
 \displaystyle{
       {\partial^2T\over\partial t^2}(r_0,t) +
      {1\over\tau}\,{\partial T\over\partial t}(r_0,t) =
      {3k\over\rho_0 c_0 r_0\tau}\,{\partial T\over\partial r}(r_0,t)
 }
    & & \qquad\forall t>0
 \label{electroflux}
\end{array}
\end{equation}
The last condition in Eqs.~(\ref{electroflux}) contains the density
and specific heat of the electrode (where $r\le r_0$),
given by $\rho_0=21500~\hbox{kg}\,\hbox{m}^{-3}$ and
$c_0=132~\hbox{J}\,\hbox{kg}^{-1}\hbox{K}^{-1}$, respectively.
This condition is based on the fact that in the interior and
on the surface of the spherical electrode it must hold
\begin{equation}
\label{eleccons}
      \left.
      \nabla\cdot\mathbf q(r,t) +
      \rho_0 c_0{\partial T\over\partial t}(r,t)
      \ \right\vert_{r\le r_0}\!
      = 0,
\end{equation}
which just states that the heat flux $\mathbf q$ spreading into the
biological tissue is due to the change of the thermal energy content
of the electrode. Applying Gauss' law to Eq.~(\ref{eleccons})
and integrating over the entire volume $r\le r_0$ yields
\begin{equation}
\label{rfhcond}
   \oint\mathbf q(r,t)\cdot\hat{\mathbf e}_r\,r_0^2\,d\Omega
   + \rho_0 c_0 {4\pi r_0^3\over3}{\partial T\over\partial t}(r,t) = 0,
\end{equation}
where we have assumed that the electrode has relatively small radius and
sufficiently high thermal conductivity, which makes the electrode act like
a punctual heat source. In the derivation of Eq.~(\ref{rfhcond}), this
justifies taking $\rho_0 c_0$ to be constant and pulling $\partial T/\partial t$
out of the volume integral.

Finally, by substituting the expression for the hyperbolic heat flux
Eq.~(\ref{hypflux}) into Eq.~(\ref{rfhcond}) and using
$\nabla T\cdot\hat{\mathbf e}_r=\partial T/\partial r$, where
$\hat{\mathbf e}_r$ is the usual radial unit vector, one readily obtains the
last condition for the electrode in Eqs.~(\ref{electroflux}).

\begin{figure}
\centering
\includegraphics[width=13cm]{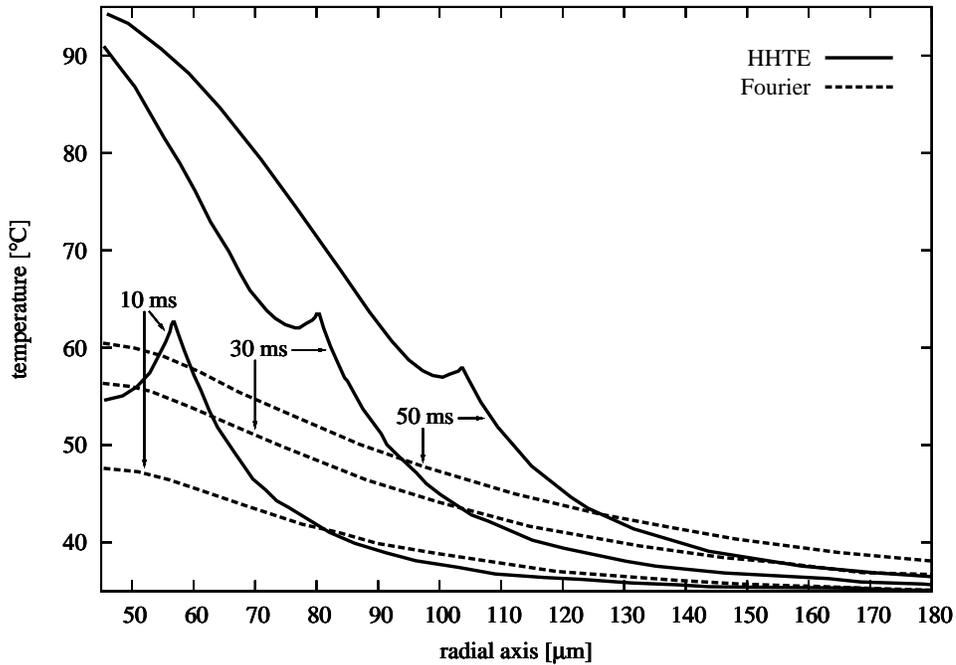}
\caption{Temperature distributions for the cornea along the radial axis
for three different measured times $t=10, 30$, and $50$~ms of a
non-pulsed {\sl RF} application with total power $P=30~\hbox{mW}$
and duration $\Delta t=600~\hbox{ms}$. The initial temperature is
$T_0=35^{\circ}\hbox{C}$ and the electrode radius $r_0=45\ \mu\hbox{m}$.
The relaxation time is taken $\tau=0.1\kern.5pt\hbox{s}$.
}
\label{rfhpic}
\end{figure}

The analytical solution of this problem and the corresponding {\sl PHTE}
solution have been obtained in Ref.~\cite{molina} as
an application for corneal {\sl RFH}.
We have visualized the data generated from Ref.~\cite{todos}
in Fig.~\ref{rfhpic} in order to discuss the main characteristics of the model.
Figure~\ref{rfhpic} represents the hyperbolic (solid lines) and parabolic
(dashed lines) temperature profiles along the radial axis for the times
$t=10, 30$ and $50$~ms. Similar to the laser-heating case, for small
times and locations near the electrode surface the temperatures from
the {\sl HHTE} are greater than from the {\sl PHTE}.
This fact also becomes important when instead of a non-pulsed {\sl RFH}
application a pulsed power is applied. Moreover, the presence of
pronounced peaks in Fig.~\ref{rfhpic} reveals the wave character of the
{\sl HHTE} model and its prediction of a finite heat conduction speed
$v=\sqrt{\alpha/\tau}\approx1.17~\hbox{mm}/\hbox{s}$.

\section{CONCLUSIONS AND OUTLOOK}
\label{conclusions}

In this paper, we have outlined the fundamentals and
main differences between the parabolic and hyperbolic model of
heat conduction. These differences encourage the use of the {\sl HHTE}
approach in processes in which great amounts of heat are transferred to
any material in very short times. Specifically, laser and radiofrequency
heating are two surgical techniques of this type of processes.
Through the application of {\sl HHTE} to these surgical techniques, we
have shown the main characteristics of this model and its
differences with {\sl PHTE} predictions.

At the moment, we are working on the analytical modeling of the
hyperbolic bioheat equation including a source term for blood perfusion
({\it viz.}\/ Eq.~(\ref{bioheat})). This will be less important for
non-perfused tissues (such as the cornea), but relevant for well-perfused
organs such as {\it e.g.}\/ the liver.

On the other hand, we are studying the implications of the hyperbolic
heat equation for the case of pulsed {\sl RF} applications. Energy
pulses for surgical procedures are being employed in such different
areas as conductive keratoplasty ({\sl CK}\/) and {\sl RFH} to destroy
tumors.

However, we are well aware of the restrictions of purely analytical models,
mainly due to the lack of mathematical tractability. For this reason,
we are planning to develop alternative theoretical models based on numerical
techniques (such as finite elements) in order to use the hyperbolic bioheat
equation in models with more complicated geometry (more realistic
electrode and tissue geometries with irregular boundaries) and also
varying thermal characteristics.

% \appendix

% graphics come at the end

\newpage
\vskip1cm

% Figure 1 - Current conservation

\vskip1cm

% Figure 2 - Fourier's law

\eject
\newpage
\vskip1cm

% Figure 3 - Laser model geometry

\vskip1cm

% Figure 4 - Laser ablation

\eject
\newpage

% Figure 5 - RF heating geometry

% Figure 6 - RFH plot

\eject

\end{document}